\documentclass[Journal,SectionNumbers,SingleSpace,InsideFigs]{ascelike}

\usepackage[utf8]{inputenc}
\usepackage[OT1]{fontenc}
\usepackage{graphicx}
\usepackage[english]{babel}

\usepackage{amsmath}
\usepackage{amsfonts}
\usepackage{amssymb}
\usepackage{amsthm}
\usepackage{bm}

\usepackage[usenames,dvipsnames]{xcolor}
\usepackage{booktabs}
\usepackage{url}
\usepackage{tikz}
\usetikzlibrary{%
   arrows,%
   calc,%
   fit,%
   patterns,%
   plotmarks,%
   shapes.geometric,%
   shapes.misc,%
   shapes.symbols,%
   shapes.arrows,%
   shapes.callouts,%
   shapes.multipart,%
   shapes.gates.logic.US,%
   shapes.gates.logic.IEC,%
   er,%
   automata,%
   backgrounds,%
   chains,%
   topaths,%
   trees,%
   petri,%
   mindmap,%
   matrix,%
   calendar,%
   folding,%
   fadings,%
   through,%
   patterns,%
   positioning,%
   scopes,%
   decorations.fractals,%
   decorations.shapes,%
   decorations.text,%
   decorations.pathmorphing,%
   decorations.pathreplacing,%
   decorations.footprints,%
   decorations.markings,%
   shadows}

\usepackage{lineno} 
\usepackage[bookmarks]{hyperref}

\newcommand{\dd}{\,\mathrm{d}}

\newcommand{\mbf}[1]{\mathbf{#1}}

\renewcommand{\vec}[1]{{\bm#1}}

\newcommand{\uz}{^{(0)}} 
\newcommand{\un}{^{(n)}} 

\newcommand{\ul}[1]{\underline{#1}}
\newcommand{\ol}[1]{\overline{#1}}

\newcommand{\Rsys}{R_\text{sys}}
\newcommand{\lRsys}{\ul{R}_\text{sys}}
\newcommand{\uRsys}{\ol{R}_\text{sys}}

\def\Rsys{R_\text{sys}}
\def\Tsys{T_\text{sys}}

\newcommand{\E}{\operatorname{E}}
\newcommand{\V}{\operatorname{Var}}
\newcommand{\wei}{\operatorname{Wei}} 
\newcommand{\ig}{\operatorname{IG}}   

\newcommand{\El}{\ul{\operatorname{E}}}
\newcommand{\Eu}{\ol{\operatorname{E}}}

\def\yz{y\uz}
\def\yn{y\un}
\newcommand{\yfun}[1]{y^{({#1})}}

\def\ykz{y\uz_k}
\def\ykn{y\un_k}

\def\yzl{\ul{y}\uz}
\def\yzu{\ol{y}\uz}

\def\ykzl{\ul{y}\uz_k}
\def\ykzu{\ol{y}\uz_k}

\newcommand{\ykzlfun}[1]{\ul{y}\uz_{#1}}
\newcommand{\ykzufun}[1]{\ol{y}\uz_{#1}}

\def\nz{n\uz}
\def\nn{n\un}
\newcommand{\nfun}[1]{n^{({#1})}}

\def\nkz{n\uz_k}
\def\nkn{n\un_k}
\newcommand{\nkzfun}[1]{n\uz_{#1}}
\newcommand{\nkzlfun}[1]{\ul{n}\uz_{#1}}
\newcommand{\nkzufun}[1]{\ol{n}\uz_{#1}}

\def\nzl{\ul{n}\uz}
\def\nzu{\ol{n}\uz}

\def\nkzl{\ul{n}\uz_k}
\def\nkzu{\ol{n}\uz_k}

\def\taut{\tau(\vec{t})}

\def\MZ{\mathcal{M}\uz}
\def\MN{\mathcal{M}\un}

\def\MkZ{\mathcal{M}\uz_k}

\def\PkZ{\Pi\uz_k}

\newcommand{\PZi}[1]{\Pi\uz_{#1}}

\def\tnow{t_\text{now}}
\def\tpnow{t^+_\text{now}}

\allowdisplaybreaks

\begin{document}
\title{ROBUST BAYESIAN RELIABILITY FOR COMPLEX SYSTEMS UNDER PRIOR-DATA CONFLICT}

\author{
Gero Walter%
\thanks{
Postdoctoral Researcher,
School of Industrial Engineering,
Eindhoven University of Technology, Eindhoven, The Netherlands.
E-mail: g.m.walter@tue.nl. Corresponding author.},
\ Ph.D.
\\
and
Frank P.A. Coolen%
\thanks{
Professor,
Department of Mathematical Sciences,
Durham University, Durham, United Kingdom.
E-mail: frank.coolen@durham.ac.uk.},
\ Ph.D.
}

\maketitle

%
%
\begin{abstract}
In reliability engineering, data about failure events is often scarce.
To arrive at meaningful estimates for the reliability of a system,
it is therefore often necessary to also include expert information in the analysis,
which is straightforward in the Bayesian approach by using an informative prior distribution.
A problem called prior-data conflict then can arise:
observed data seem very surprising from the viewpoint of the prior,
i.e., information from data is in conflict with prior assumptions.
Models based on conjugate priors can be insensitive to prior-data conflict,
in the sense that the spread of the posterior distribution does not increase in case of such a conflict,
thus conveying a false sense of certainty.
An approach to mitigate this issue is presented, by considering sets of prior distributions
to model limited knowledge on Weibull distributed component lifetimes,
treating systems with arbitrary layout using the survival signature.
This approach can be seen as a robust Bayesian procedure or imprecise probability method
that reflects surprisingly early or late component failures
by wider system reliability bounds.
%
\end{abstract}

\KeyWords{System Reliability, Imprecise Probability, Survival Signature, Robust Bayesian Methods, Remaining Useful Life}

\section{Introduction}
\label{sec:intro}
In reliability engineering, a central task is to describe the reliability of a complex system.
This is usually done by determining the \emph{reliability function} $R(t)$,
in other contexts also known as the \emph{survival function},
giving the probability that the system has not failed by time $t$:
\begin{linenomath*}
\begin{align}
\Rsys(t) = P(\Tsys \geq t)\,,
\end{align}
\end{linenomath*}
where $\Tsys$ is the random variable giving the failure time of the system. %
Based on the distribution of $\Tsys$, which can also be expressed
in terms of the cumulative distribution function $F_\text{sys}(t) = 1 - \Rsys(t)$,
decisions about, e.g., scheduling of maintenance work can be made.

Often, there is no failure data for the system itself
(e.g., if the system is a prototype, or the system is used under unique circumstances),
but there is some information about failure times for the components the system is made of.
The proposed method allows to analyse systems of arbitrary system structures, 
i.e., any combination and nesting of series, parallel, k-out-of-n, or bridge-type arrangements,
by use of the survival signature \cite{2012:survsign}.
In doing so, the present paper extends similar previous work
which focused on a simple parallel system with homogeneous components \cite{2015:walter}.
In this paper, we assume that components can be divided into $K$ different groups,
and components within each group $k$ ($k=1, \ldots, K$) can be assumed to be exchangeable,
i.e., to follow the same failure time distribution.
Components of group $k$ are denoted as \emph{type $k$ components},
and are assumed to be independent from components of other types.
Type $k$ component lifetimes $T_i^k$ ($i = 1, \ldots, n_k$)
are assumed as Weibull distributed with type-specific parameters,
where the shape parameter $\beta_k$ is known
and the scale parameter $\lambda_k$ is unknown.

Focusing on a running system,
we assume that observations consist solely of the failure times of components in this system up to time $\tnow$,
such that the failure times of components that have not failed by $\tnow$ are right-censored,
and calculate $\Rsys(t \mid t > \tnow) = P(\Tsys \geq t \mid t > \tnow)$,
which can be used to determine the remaining useful life of the system
(in short RUL, see, e.g., \citeNP{2014:rul-review}).

The Bayesian approach allows to base estimation of the component failure distributions
on both data and further knowledge not given by the data,
the latter usually provided in the form of expert knowledge.
This knowledge is encoded in form of a so-called prior distribution,
which here, as the shape parameter $\beta_k$ is assumed to be known,
is a distribution on the scale parameter $\lambda_k$. 
Expert knowledge is especially important when there is very few data on the components, 
as only with its help meaningful estimates for the system reliability can be made.
Due to the iterative nature of the Bayesian framework,
it is also possible to use a set of posteriors based on component test data as the set of priors
instead of a purely expert-based set of priors as discussed so far.
The component test data may also contain right-censored observations,
which can be treated in the same way as those from the running system
(see Section~\ref{sec:lambdawithcens} for details).

The choice of prior distribution to encode given expert knowledge is often debatable,
and a specific choice of prior is difficult to justify.
A way to deal with this is to employ sensitivity analysis,
i.e., studying the effect of different choices of prior distribution on the quantities of interest
(in our case, the system reliability function, which, in Bayesian terms, is a posterior predictive distribution).
This idea has been explored in systematic sensitivity analysis, or robust Bayesian methods
(for an overview on this approach, see, e.g.,
\citeNP{1994:berger} or \citeNP{2000:rios}). 

The work presented here can be seen as belonging to the robust Bayesian approach
since it uses sets of priors. However, our focus and interpretation is slightly different,
as we consider the result of our procedure, sets of reliability functions, as the proper result,
while a robust Bayesian would base his analyses on a single reliability function from the set
in case (s)he was able to conclude that quantities of interest are not `too sensitive' to the choice of prior.
In contrast, our viewpoint is rooted in the theory of imprecise or interval probability \cite{1991:walley,itip},
where sets of distributions are used to express the precision of probability statements themselves:
the smaller the set, the more precise the probability statement.
Indeed, the system reliability function $\Rsys(t)$ is a collection of probability statements,
and a small set for $\Rsys(t)$ will indicate that the reliability of the system can be can quantified quite precisely,
while a large set will indicate that available knowledge about $\Tsys$ is rather shaky.
In line with imprecise or interval probability methods, the method provides, for each $t$,
a lower reliability $\lRsys(t) = \ul{P}(T_\text{sys} \geq t)$,
and an upper reliability $\uRsys(t) = \ol{P}(T_\text{sys} \geq t)$.
Sections~\ref{sec:modforsurpr} and \ref{sec:robrel} will explain how these bounds are obtained
based on sets of prior distributions on the scale parameters of the component lifetime distributions.

The central merit of the proposed method is that it adequately reflects prior-data conflict
(see, e.g., \citeNP{2006:evans}),
i.e.\ the conflict that can arise between prior assumptions on component lifetimes
and observed behaviour of components in the system under study.
As will be shown in Section~\ref{sec:weibull}, when taking the standard choice of a conjugate prior,
prior-data conflict is ignored, as the spread of the posterior distribution does not increase in case of such a conflict,
ultimately conveying a false sense of certainty
by communicating that the reliability of a system can be quantified quite precisely when in fact it can not.
In contrast, the proposed method will indicate prior-data conflict by wider bounds for $\Rsys(t)$.
This behaviour is obtained by a specific choice for the set of priors (see \citeNP{Walter2009a} and \citeNP{diss} \S 3.1.4)
which leads to larger sets of posterior distributions when prior knowledge and data are in conflict
(see Section~\ref{sec:modforsurpr} for more details).
If a prior based on component test data is used,
such prior-data conflict sensitivity furthermore allows to uncover a conflict between
current observations in the running system and past observations from component tests.

The paper is organized as follows.
In Section~\ref{sec:weibull}, a Bayesian analysis for Weibull component lifetimes
with fixed shape parameter is described,
illustrating the issue of prior-data conflict.
Section~\ref{sec:modforsurpr} then details the use of sets of priors
for the scale parameter $\lambda_k$ of the Weibull distribution,
showing how this mitigates the prior-data conflict issue for $\lambda_k$.
Section~\ref{sec:robrel} then develops a method to calculate system reliability bounds
for a running system based on sets of priors for $\lambda_k$. 
Section~\ref{sec:elicitation} discusses elicitation,
by giving guidelines on how to choose prior parameter bounds
that reflect expert information on components.
Section~\ref{sec:examples} contains examples illustrating the merits of our method,
by studying the effect of surprisingly early or late component failures,
showing that observations in conflict to prior assumptions
indeed lead to more cautious system reliability predictions.
Section~\ref{sec:concluding} concludes the paper by summarizing results
and indicating avenues for further research.

\section{Bayesian Analysis of Weibull Lifetimes}
\label{sec:weibull}

Consider a system with components of $k=1,\ldots,K$ different types;
for each type $k$, there are $n_k$ exchangeable components in the system.
For each type $k$ component, we assume for its lifetime $T_i^k$ ($i=1,\ldots,n_k$, $k = 1, \ldots, K$)
a Weibull distribution with fixed shape parameter $\beta_k > 0$,
in short $T_i^k \mid \lambda_k \sim \wei(\beta_k,\lambda_k)$,
with density and cdf%
\begin{linenomath*}
\begin{align}
\label{eq:weibulldens}
f(t_i^k \mid \lambda_k) &= \frac{\beta_k}{\lambda_k} (t_i^k)^{\beta_k-1} e^{-\frac{(t_i^k)^{\beta_k}}{\lambda_k}}\,, \\
\label{eq:weibullcdf}
F(t_i^k \mid \lambda_k) &= 1 - e^{-\frac{(t_i^k)^{\beta_k}}{\lambda_k}} = P(T_i^k \leq t_i^k \mid \lambda_k)\,,
\end{align}
\end{linenomath*}
where $\lambda_k > 0$ and $t > 0$.

The shape parameter $\beta_k$ determines whether the hazard rate is increasing ($\beta_k > 1$)
or decreasing ($\beta_k < 1$) over time.
For $\beta_k=1$, one obtains the Exponential distribution with constant hazard rate as a special case.
The value for $\beta_k$ will often be clear from practical considerations.

The scale parameter $\lambda_k$ can be interpreted through the relation
\begin{linenomath*}
\begin{align}
\E[T_i^k \mid \lambda_k] &= \lambda_k^{1/\beta_k}\, \Gamma(1 + 1/\beta_k)\,.
\label{eq:lambdainterpret}
\end{align}
\end{linenomath*}
For encoding expert knowledge about the reliability of the components,
one needs to assign a prior distribution over the scale parameter $\lambda_k$.
A convenient choice is to use the inverse Gamma distribution,
commonly parametrized in terms of the hyperparameters $a_k > 0$ and $b_k > 0$:
\begin{linenomath*}
\begin{align}
f(\lambda_k\mid a_k,b_k) &= \frac{(b_k)^{a_k}}{\Gamma(a_k)} \lambda_k^{-a_k -1} e^{-\frac{b_k}{\lambda_k}}
\label{eq:ig-def}
\end{align}
\end{linenomath*}
in short, $\lambda_k \mid a_k, b_k \sim \ig(a_k,b_k)$.
The inverse Gamma is convenient because it is a conjugate prior,
i.e., the posterior obtained by Bayes' rule is again inverse Gamma and thus easily tractable;
the prior parameters only need to be updated to obtain the posterior parameters.

In the standard Bayesian approach, 
one has to fix a prior by choosing values for $a_k$ and $b_k$
to encode specific prior information about component lifetimes.
In our imprecise approach, we allow instead these parameters
to vary in a set, this is advantageous also
because expert knowledge is often vague,
and it is difficult for the expert(s) to pin down precise hyperparameter values.
For the definition of the hyperparameter set,
we use however a parametrization in terms of $\nz > 1$ and $\yz > 0$ instead of $a_k$ and $b_k$,
where we drop the index $k$ for the discussion about the prior model in the following,
keeping in mind that each component type will have its own specific parameters.
We use
$\nz = a - 1$ and
$\yz = b / \nz$,
where $\yz$ can be interpreted as the prior guess for the scale parameter $\lambda$,
as $\E[\lambda\mid\nz,\yz] = \yz$.
This parametrization also makes the nature of the combination
of prior information and data through Bayes' rule more clear:
After observing $n$ component lifetimes $\vec{t} = (t_1, \ldots, t_n)$,
the updated parameters are
\begin{linenomath*}
\begin{align}
\nn &= \nz + n\,, 
&
\yn &=  \frac{\nz \yz + \taut}{\nz + n}\,,
\label{eq:ig-update}
\end{align}
\end{linenomath*}
where $\taut = \sum_{i=1}^n (t_i)^\beta$. %
We thus have
\begin{linenomath*}
\begin{align}
\lambda \mid \nz, \yz, \vec{t} \sim \ig(\nz + n + 1, \nz \yz + \taut). 
\label{eq:ig-update-alpha}
\end{align}
\end{linenomath*}
From the simple update rule \eqref{eq:ig-update}, we see that
$\yn$ is a weighted average of the prior parameter $\yz$ and the maximum likelihood (ML) estimator $\taut/n$,
with weights proportional to $\nz$ and $n$, respectively.
$\nz$ can thus be interpreted as a prior strength or pseudocount,
indicating how much our prior guess should weigh against the $n$ observations.
Furthermore, $\V[\lambda\mid\nz,\yz] = (\yz)^2 / (1 - 1/\nz)$,
so for fixed $\yz$, the higher $\nz$,
the more probability mass is concentrated around $\yz$. 

\begin{figure}
\centering
\includegraphics[width=0.8\textwidth]{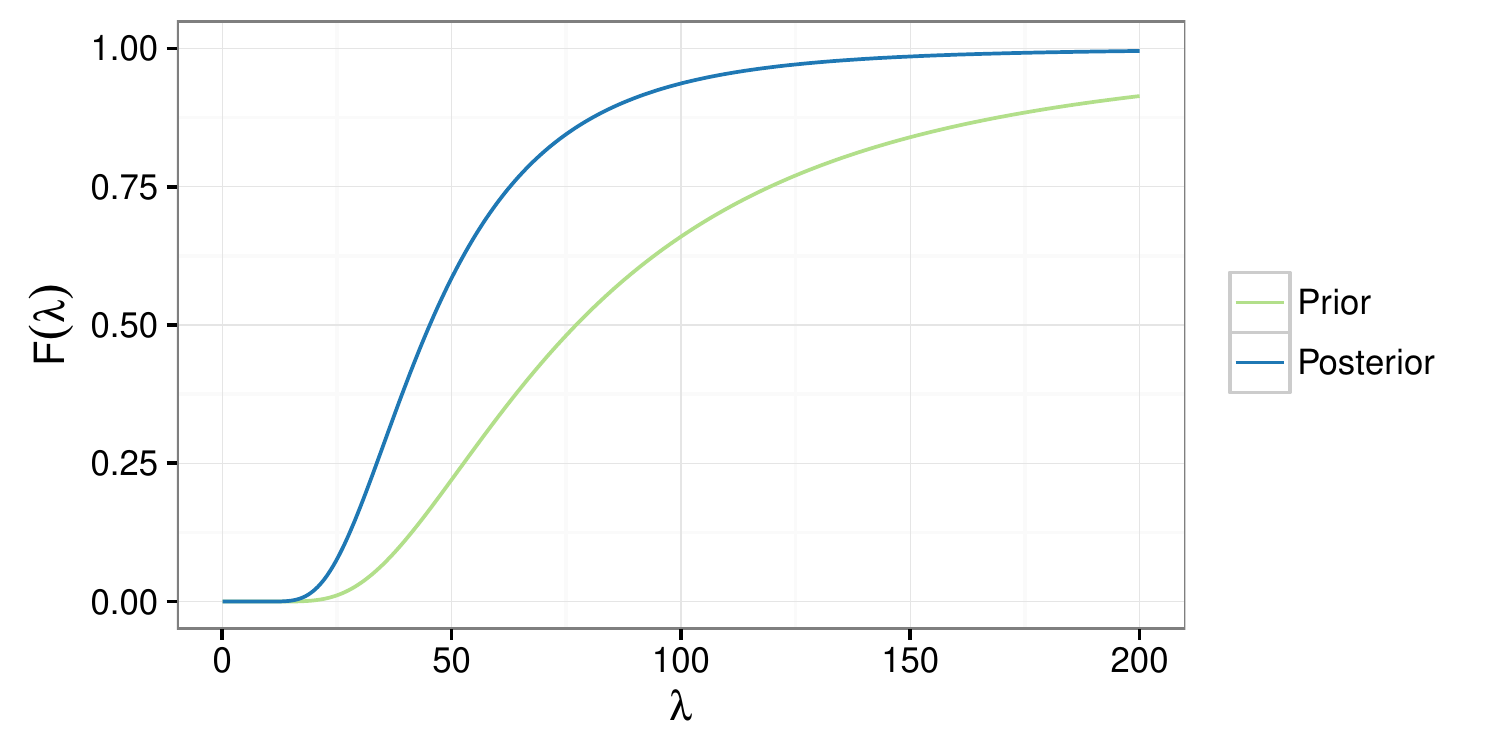}
\caption{Prior and posterior cdf for $\lambda$ given surprising observations;
the conflict between prior assumptions and data is averaged out,
with a more pointed posterior giving a false sense of certainty.}
\label{fig:weibull-pdc}
\end{figure}

However, the weighted average structure for $\yn$
is behind the problematic behaviour in case of prior-data conflict.
Assume that from expert knowledge we expect
to have a mean component lifetime of 9 weeks.
Using \eqref{eq:lambdainterpret}, with $\beta=2$ we obtain $\yz = 103.13$.
We choose $\nz = 2$, so our prior guess for the mean component lifetime
counts like having two observations with this mean.
If we now have a sample of two observations
with surprisingly early failure times $t_1 = 1$ and $t_2 = 2$,
using \eqref{eq:ig-update} we get $\nfun{2} = 4$
and $\yfun{2} = \frac{1}{4}(2 \cdot 103.13 + 1^2 + 2^2) = 52.82$,
so our posterior expectation for the scale parameter $\lambda$ is $52.82$,
equivalent to a mean component lifetime of $6.44$ weeks.
The posterior standard deviation (sd) for $\lambda$ is $60.99$.
Compared to the prior standard deviation of $145.85$,
the posterior expresses now more confidence that mean lifetimes are around $\yfun{2} = 52.82$
than the prior had about $\yz = 103.13$.
This irritating conclusion is illustrated in Figure~\ref{fig:weibull-pdc};
the posterior cdf is shifted halfway towards the values for $\lambda$
that the two observations suggest
(the ML estimator for $\lambda$ would be $2.5$),
and is steeper than the prior (so the pdf is more pointed),
thus conveying a false sense of certainty about $\lambda$.
We would obtain almost the same 
posterior distribution
if we had assumed the mean component lifetime to be 7 weeks (so $\yz = 62.39$),
and observed lifetimes $t_1 = 6$, $t_2 = 7$ in line with our expectations.
It seems unreasonable to make the same probability statements on component lifetimes in these two fundamentally different scenarios.

Note that this is a general problem in Bayesian analysis with canonical conjugate priors.
For such priors, the same update formula \eqref{eq:ig-update} applies,
and so conflict is averaged out, for details see \citeN{Walter2009a} and \citeN[\S 3.1.4 and \S A.1.2]{diss}.
%

\section{Models reflecting surprising data}
\label{sec:modforsurpr}

Despite this issue of ignoring prior-data conflict,
the tractability of the update step
is a very attractive feature of the conjugate setting.
As was shown by \citeN{Walter2009a} (see also \citeNP{diss}, \S 3.1),
it is possible to retain tractability and to have a meaningful reaction to prior-data conflict
when using sets of priors generated by varying both $\nz$ and $\yz$.
Then, the magnitude of the set of posteriors,
and with it the precision of posterior probability statements,
will be sensitive to the degree of prior-data conflict,
i.e.\ leading to more cautious probability statements when prior-data conflict occurs.

Instead of a single prior guess $\yz$ for the mean component lifetimes,
we will now assume a range of prior guesses $[\yzl, \yzu]$, and also a range $[\nzl, \nzu]$ of pseudocounts,
i.e., we now consider the set of priors
\begin{linenomath*}
\begin{align}
\MZ := \{ f(\lambda\mid\nz,\yz) \mid \nz \in [\nzl, \nzu], \yz \in [\yzl, \yzu] \}
\label{eq:setofpriors}
\end{align}
\end{linenomath*}
to express our prior knowledge about component lifetimes.
Each of the priors $f(\lambda\mid\nz,\yz)$ is then updated to the posterior
$f(\lambda\mid\nz,\yz,\vec{t}) = f(\lambda\mid\nn,\yn)$
by using \eqref{eq:ig-update},
such that the set of posteriors $\MN$ can be written as
$\MN = \{ f(\lambda\mid\nn,\yn) \mid \nz \in [\nzl, \nzu], \yz \in [\yzl, \yzu] \}$.
This procedure of using Bayes' Rule element by element
is seen as self-evident in the robust Bayesian literature,
but can be formally justified as being \emph{coherent}
(a self-consistency property)
in the framework of imprecise probability, where it is known as
\emph{Generalized Bayes' Rule} \cite[\S 6.4]{1991:walley}.

Technically, it is crucial to consider a range of pseudocounts $[\nzl, \nzu]$
along with the range of prior guesses $[\yzl, \yzu]$,
as only then $\taut/n \not\in [\yzl, \yzu]$
leads to the set of posteriors being larger
and hence reflecting prior-data conflict.

\begin{figure}
\includegraphics[width=\textwidth]{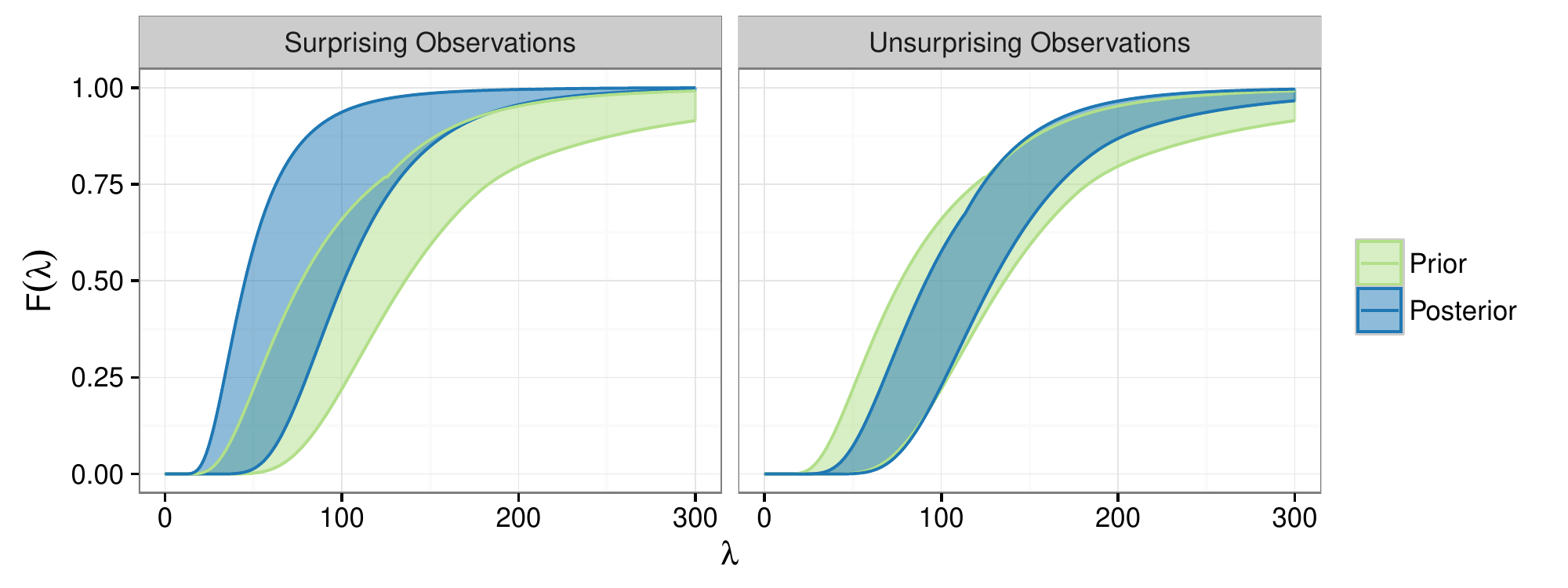}
\caption{Set of prior and posterior cdfs for $\lambda$ for two surprising observations $t_1 = 1$, $t_2 = 2$ (left)
and two unsurprising observations $t_1 = 10$, $t_2 = 11$ (right).}
\label{fig:setofpost-pdc-nopdc}
\end{figure}

Continuing the example from Section~\ref{sec:weibull} and Figure~\ref{fig:weibull-pdc},
assume now for the mean component lifetimes the range 9 to 11 weeks,
this corresponds to $[\yzl,\yzu] = [103.13, 154.06]$.
Choosing $[\nzl,\nzu] =[2, 5]$
means to value this information on mean component lifetimes as equivalent to having seen two to five observations.
Compare now the set of posteriors obtained from observing
$t_1 = 1$, $t_2 = 2$ (as before), see Figure~\ref{fig:setofpost-pdc-nopdc} (left),
and $t_1 = 10$, $t_2 = 11$, see Figure~\ref{fig:setofpost-pdc-nopdc} (right).
There is now a clear difference between the two scenarios of
observations in line with expectations and observations in conflict.
In the prior-data conflict case, the set of posteriors (blue)
is shifted towards the left, but has about the same size as the set of priors (yellow),
and so posterior quantification of reliability has the same precision,
despite having seen two observations.
Instead, in the no conflict case, the set of posteriors is smaller than the set of priors,
such that the two observations have increased the precision of reliability statements.

As each posterior in $\MN$ corresponds to a predictive distribution for $\Tsys$,
we will have a set of reliability functions $\Rsys(t)$.
The derivation of $\Rsys(t)$ for systems with $K$ component types and arbitrary layout
will be given in Section~\ref{sec:robrel} below.
This will include, in contrast to previous studies using sets of priors of type 
\eqref{eq:setofpriors}, the treatment of censored observations.
More specifically, we consider the case of \emph{non-informative right censoring},
where the censoring process is independent of the failure process.

\newpage
\section{Robust Reliability for Complex Systems via the Survival Signature}
\label{sec:robrel}

Consider now a system of arbitrary layout,
consisting of components of $K$ types.
This system is observed until time $\tnow$,
leading to censored observation of lifetimes of components within the system,
and we will explain in Section~\ref{sec:lambdawithcens}
how the scale parameter $\lambda_k$ for component type $k$
can be estimated in this situation.
Section~\ref{sec:sysrelwithsurvsign} describes then
how the system reliability function can be efficiently obtained using the survival signature.
For this, we need the posterior predictive distribution
of the number of components that function at times $t > \tnow$,
which we will derive in Section~\ref{sec:postpred}.
Finally, in Section~\ref{sec:optimize} we describe how
the lower and upper bound for the system reliability function
are obtained when prior parameters $(\nkz,\ykz)$
vary in sets $\PkZ = [\nkzl,\nkzu] \times [\ykzl,\ykzu]$,
defining sets $\MkZ$ of prior distributions over $\lambda_k$,
as introduced in Section~\ref{sec:modforsurpr}.

\subsection{Bayesian Estimation of Component Scale Parameter with Right-censored Lifetimes}
\label{sec:lambdawithcens}

Consider observing a system until $\tnow$,
where the system has $K$ different types of components,
and for each type $k$ there are $n_k$ components in the system.
Denoting the number of type $k$ components that have failed by $\tnow$ by $e_k$,
there are $n_k - e_k$ components still functioning at $\tnow$.
We denote the corresponding vector of observations by
\begin{linenomath*}
\begin{align}
\vec{t}^k_{e_k;n_k} &= \big( \underbrace{t^k_1, \ldots, t^k_{e_k}}_{e_k \text{failure times}},
                             \underbrace{\tpnow, \ldots, \tpnow}_{n_k-e_k \text{censored obs.}} \big)\,,
\end{align}
\end{linenomath*}
where $t^+$ indicates a right-censored observation.
%

According to Bayes' rule, multiplying the prior density and the likelihood
(which accounts for right-censored observations through the cdf terms)
gives a term proportional to the density of the posterior distribution for $\lambda_k$:
\begin{linenomath*}
\begin{align}
f(\lambda_k\mid\nkz,\ykz,\vec{t}^k_{e_k;n_k})
 &\propto f(\lambda_k)
          \big[ 1- F(\tnow\mid\lambda_k) \big]^{n_k-e_k}
          \prod_{i=1}^{e_k} f(t_i^k \mid \lambda_k) 
\end{align}
\end{linenomath*}
Conjugacy is preserved and we get 
$\lambda_k\mid\nkz,\ykz,\vec{t}^k_{e_k;n_k} \sim \ig(\nkn + 1, \nkn\ykn)$, where
\begin{linenomath*}
\begin{align}
\nkn + 1 &= \nkz + e_k + 1 \\
\nkn\ykn &= \nkz\ykz + (n_k-e_k) (\tnow)^\beta + \sum_{i=1}^{e_k} (t_i^k)^\beta
\end{align}
\end{linenomath*}
are the updated parameters of the inverse gamma distribution.

\subsection{System Reliability using the Survival Signature}
\label{sec:sysrelwithsurvsign}

The structure of complex systems can be visualized by reliability block diagrams,
an example is given in Figure~\ref{fig:brakesys-layout}.
Components are represented by boxes or nodes,
and the system works when a path from the left end to the right exists
which passes only through working components.
In a system with $n$ components, the state of the components can be expressed by the state vector
$\vec{x} = (x_1,x_2,\ldots,x_n) \in \{0,1\}^n$, with $x_i=1$ if the $i$th component functions 
and $x_i=0$ if not.
The structure function $\phi : \{0,1\}^n \rightarrow \{0,1\}$, defined for all possible $\vec{x}$, takes 
the value 1 if the system functions and 0 if the system does not function for state vector $\vec{x}$
\cite{BP75}.
Most real-life systems are coherent,
which means that $\phi(\vec{x})$ is non-decreasing in any of the components of $\vec{x}$,
so system functioning cannot be improved by worse performance of one or more of its components.
Furthermore, one can usually assume that $\phi(0, \ldots, 0) = 0$ and $\phi(1, \ldots, 1) = 1$.

The survival signature \cite{2012:survsign} is a summary of the structure function
for systems with $K$ groups of exchangeable components.
Denoted by $\Phi(l_1,\ldots,l_K)$, with $l_k=0,1,\ldots,n_k$ for $k=1,\ldots,K$,
it is defined as the probability for the event that the system functions
given that precisely $l_k$ of its $n_k$ components of type $k$ function, for each $k\in \{1,\ldots,K\}$.
Essentially, this creates a $K$-dimensional partition for the event $\Tsys > t$,
such that $\Rsys(t) = P(\Tsys > t)$ can be calculated using the law of total probability,
\begin{linenomath*}
\begin{align}
P(\Tsys > t) &= \sum_{l_1=0}^{m_1} \cdots \sum_{l_K=0}^{m_K} P(\Tsys > t \mid C^1_t = l_1,\ldots, C^K_t = l_K)
                                                                                  P\Big( \bigcap_{k=1}^K \{ C^k_t = l_k\} \Big) \nonumber\\
             &= \sum_{l_1=0}^{n_1} \cdots \sum_{l_K=0}^{n_K} \Phi(l_1,\ldots,l_K) P\Big( \bigcap_{k=1}^K \{ C^k_t = l_k\} \Big) \nonumber\\
             &= \sum_{l_1=0}^{n_1} \cdots \sum_{l_K=0}^{n_K} \Phi(l_1,\ldots,l_K) \prod_{k=1}^K P(C^k_t = l_k)\,,
\label{eq:sysrel-survsign}
\end{align}
\end{linenomath*}
where $P(C^k_t = l_k)$ is the (predictive) probability that exactly $l_k$ components of type $k$ function at time $t$,
and the last equality holds as we assume that components of different types are independent.
Note that for coherent systems,
the survival signature $\Phi(l_1,\ldots,l_K)$ is non-decreasing in each $l_k$.
%

\subsection{Posterior Predictive Distribution}
\label{sec:postpred}

In calculating the system reliability using \eqref{eq:sysrel-survsign},
the component-specific predictive probabilities $P(C^k_t = l_k)$
need to use all information available at time $\tnow$,
which, in the Bayesian framework,
are given by the posterior predictive distribution 
$P(C^k_t = l_k\mid\nkz,\ykz, \vec{t}^k_{e_k;n_k})$, $l_k = 0, 1, \ldots, n_k-e_k$. 
(Remember that $e_k$ type $k$ components have failed by $\tnow$,
such that there can be at most $n_k-e_k$ working components beyond time $\tnow$.)
This posterior predictive distribution is obtained as
\begin{linenomath*}
\begin{align}
\lefteqn{%
P(C^k_t = l_k\mid\nkz,\ykz, \vec{t}^k_{e_k;n_k})}\hspace*{5ex} \nonumber\\  %
 &= { n_k - e_k \choose l_k} \int \big[P(T^k >    t \mid T^k > \tnow, \lambda_k)\big]^{l_k} \times \nonumber\\ & \hspace*{17ex}
                                  \big[P(T^k \leq t \mid T^k > \tnow, \lambda_k)\big]^{n_k - e_k - l_k}
    f(\lambda_k\mid\nkz,\ykz,\vec{t}^k_{e_k;n_k}) \dd \lambda_k\,.
\label{eq:postpredtnow}
\end{align}
\end{linenomath*}
Now, by the Weibull assumption \eqref{eq:weibullcdf}, one has
\begin{linenomath*}
\begin{align}
P(T^k \leq t \mid T^k > \tnow, \lambda_k)
 &= \frac{P(\tnow < T^k \leq t \mid\lambda_k)}{P(T^k > \tnow \mid \lambda_k)} \nonumber\\
 &= \frac{F(t\mid\lambda_k) - F(\tnow\mid\lambda_k)}{1-F(\tnow\mid\lambda_k)} 
  = 1 - e^{-\frac{t^{\beta_k} - (\tnow)^{\beta_k}}{\lambda_k}}\,.
\end{align}
\end{linenomath*}
With this and the posterior \eqref{eq:ig-update-alpha} substituted into \eqref{eq:postpredtnow}, this gives
\begin{linenomath*}
\begin{align}
\lefteqn{P(C^k_t = l_k\mid\nkz,\ykz, \vec{t}^k_{e_k;n_k})}\hspace*{5ex} \nonumber\\
 &= { n_k - e_k \choose l_k} \int \Big[    e^{-\frac{t^{\beta_k} - (\tnow)^{\beta_k}}{\lambda_k}}\Big]^{l_k}
                                  \Big[1 - e^{-\frac{t^{\beta_k} - (\tnow)^{\beta_k}}{\lambda_k}}\Big]^{n_k - e_k - l_k}
    \times \nonumber\\ & \hspace*{27ex}
    \frac{\big(\nkn\ykn\big)^{\nkn + 1}}{\Gamma(\nkn+1)} \lambda_k^{-(\nkn + 1) - 1} e^{-\frac{\nkn\ykn}{\lambda_k}} \dd \lambda_k \nonumber\\
 &= { n_k - e_k \choose l_k} \sum_{j=0}^{n_k-e_k-l_k} (-1)^j { n_k - e_k - l_k \choose j} \frac{\big(\nkn\ykn\big)^{\nkn + 1}}{\Gamma(\nkn+1)} 
    \times \nonumber\\ & \hspace*{13ex}
    \int \lambda_k^{-(\nkn + 1) - 1} \exp\Big\{-\frac{(l_k + j) (t^{\beta_k} - (\tnow)^{\beta_k}) + \nkn\ykn}{\lambda_k}\Big\} \dd \lambda_k\,.
\end{align}
\end{linenomath*}
The terms remaining under the integral form the core of an inverse gamma distribution \eqref{eq:ig-def}
with parameters $\nkn + 1$ and $\nkn\ykn + (l_k + j) (t^{\beta_k} - (\tnow)^{\beta_k}))$,
allowing to solve the integral using the corresponding normalization constant.
We thus have, for $l_k \in \{0,1,\ldots,n_k-e_k\}$,
\begin{linenomath*}
\begin{align}
\lefteqn{P(C^k_t = l_k\mid\nkz,\ykz, \vec{t}^k_{e_k;n_k})} \nonumber\\
 &= { n_k - e_k \choose l_k} \sum_{j=0}^{n_k-e_k-l_k} (-1)^j { n_k - e_k - l_k \choose j}
    \left(\frac{\nkn\ykn}{\nkn\ykn + (l_k + j) \big(t^{\beta_k} - (\tnow)^{\beta_k}\big)}\right)^{\nkn + 1} \nonumber\\
 &= \sum_{j=0}^{n_k-e_k-l_k} (-1)^j \frac{(n_k - e_k)!}{l_k! j! (n_k - e_k - l_k - j)!}   
    \left(\frac{\nkn\ykn}{\nkn\ykn + (l_k + j) \big(t^{\beta_k} - (\tnow)^{\beta_k}\big)}\right)^{\nkn + 1} \nonumber\\
 &= \sum_{j=0}^{n_k-e_k-l_k} (-1)^j \frac{(n_k - e_k)!}{l_k! j! (n_k - e_k - l_k - j)!} \times \nonumber\\ & \hspace*{10ex}  
    \left(\frac{\nkz\ykz + \sum_{i=1}^{e_k} (t_i^k)^{\beta_k} + (n_k-e_k)       (\tnow)^{\beta_k} }%
               {\nkz\ykz + \sum_{i=1}^{e_k} (t_i^k)^{\beta_k} + (n_k-e_k-l_k-j) (\tnow)^{\beta_k} + (l_k + j) t^{\beta_k} }\right)^{%
    \nkz + e_k + 1}.
\label{eq:postpred-priorparams}
\end{align}
\end{linenomath*}
These posterior predictive probabilities can also be expressed as a cumulative probability mass function (cmf) 
\begin{linenomath*}
\begin{align}
F(l_k \mid \nkz,\ykz,\vec{t}^k_{e_k;n_k}) = P(C^k_t \leq l_k \mid \nkz,\ykz,\vec{t}^k_{e_k;n_k}) 
 = \sum_{j=0}^{l_k} P(C^k_t = j \mid \nkz,\ykz,\vec{t}^k_{e_k;n_k})\,.
\end{align}
\end{linenomath*}

\subsection{Optimizing over Sets of Parameters}
\label{sec:optimize}

Together with \eqref{eq:postpred-priorparams},
\eqref{eq:sysrel-survsign} allows to calculate the system reliability $\Rsys(t\mid t>\tnow)$
for fixed prior parameters $(\nkz, \ykz)$, $k=1, \ldots, K$.
In Section~\ref{sec:modforsurpr}, we argued for using sets of priors $\MZ$,
which allow for vague and incomplete prior knowledge, and provide prior-data conflict sensitivity.
We will thus use, for each component type,
a set of priors $\MkZ$ defined by varying $(\nkz,\ykz)$ in a prior parameter set $\PkZ = [\nkzl,\nkzu] \times [\ykzl,\ykzu]$,
and the objective is to obtain the bounds
\begin{linenomath*}
\begin{align}
\lRsys(t \mid t > \tnow) &= \min_{\PZi{1},\ldots,\PZi{K}} \Rsys\big(t \mid t > \tnow, \cup_{k=1}^K \{\PkZ, \vec{t}^k_{e_k;n_k}\}\big)\,,
\label{eq:lrsysdef}\\
\uRsys(t \mid t > \tnow) &= \max_{\PZi{1},\ldots,\PZi{K}} \Rsys\big(t \mid t > \tnow, \cup_{k=1}^K \{\PkZ, \vec{t}^k_{e_k;n_k}\}\big)\,,
\label{eq:ursysdef}
\end{align}
\end{linenomath*}
where we suppress in notation that $\lRsys(t \mid t > \tnow)$ and $\uRsys(t \mid t > \tnow)$
depend on prior parameter sets and data.

\eqref{eq:lrsysdef} and \eqref{eq:ursysdef} seem to suggest that
a full $2K$-dimensional box-constraint optimization is necessary,
but this is not the case.
Remember that $\Phi(l_1,\ldots,l_k)$ from \eqref{eq:sysrel-survsign} is non-decreasing in each of its arguments $l_1,\ldots,l_K$,
so if there is stochastic dominance in $F(l_k \mid \nkz,\ykz,\vec{t}^k_{e_k;n_k})$,
then there is, for each component type $k$,
a prior parameter pair in $\PkZ$ that minimizes system reliability, and
a prior parameter pair in $\PkZ$ that maximizes system reliability,
independently of the other component types. 
Indeed, stochastic dominance in $F(l_k \mid \nkz,\ykz,\vec{t}^k_{e_k;n_k})$ is provided for $\ykz$.
To see this, note that $\ykz$ gives the mean expected lifetime for type $k$ components.
Thus, higher values for $\ykz$ mean higher expected lifetimes for the components,
which in turn increases the probability that many components survive until time $t$,
and with it, decreases the propability of few or no components surviving,
so in total giving low probability weight for low values of $l_k$,
and high probability weight for high values of $l_k$. 
Therefore, for any fixed value of $\nkz$, 
the lower bound of $F(l_k \mid \nkz,\ykz,\vec{t}^k_{e_k;n_k})$ for all $l_k$ is obtained with $\ykzu$, and
the upper bound of $F(l_k \mid \nkz,\ykz,\vec{t}^k_{e_k;n_k})$ for all $l_k$ is obtained with $\ykzl$.
%
%
There is however no corresponding result for $\nkz$,
such that different values of $\nkz$ may minimize (or maximize) $F(l_k \mid \nkz,\ykz,\vec{t}^k_{e_k;n_k})$ at different $l_k$'s.
Therefore, the $\nkz$ values for lower and upper system reliability bounds are obtained by numeric optimization.
%

Writing out \eqref{eq:sysrel-survsign}, one obtains
\begin{linenomath*}
\begin{align}
\lRsys(t \mid t > \tnow)
 &= \min_{\PZi{1},\ldots,\PZi{K}} \Rsys\big(t \mid t > \tnow, \cup_{k=1}^K \{\PkZ, \vec{t}^k_{e_k;n_k}\}\big) \nonumber\\
 &= \min_{\substack{\nkzfun{1} \in \left[\nkzlfun{1}, \nkzufun{1}\right]\\ \vdots\\ \nkzfun{K} \in \left[\nkzlfun{K}, \nkzufun{K}\right]}}
    \sum_{l_1=0}^{n_1-e_1} \cdots \sum_{l_K=0}^{n_K-e_K} \Phi(l_1,\ldots,l_K) \prod_{k=1}^K P(C^k_t = l_k\mid\nkz,\ykzl, \mbf{t}^k_{e_k;n_k})\,,
\label{eq:sysrel-optim-n0}
\end{align}
\end{linenomath*}
such that a $K$-dimensional box-constraint optimization is needed to obtain $\lRsys(t \mid t > \tnow)$.
The result for $\uRsys(t \mid t > \tnow)$ is completely analogous.
Computing time can furthermore be saved
by computing only those 
summation terms for which $\Phi(l_1,\ldots,l_K) > 0$.

We have implemented the method in the statistical computing environment \textsf{R} \cite{R},
using box-constraint optimization via option \texttt{L-BFGS-B} of \texttt{optim},
and intend to release a package containing functions and code to reproduce all results and figures
for the examples in Section~\ref{sec:examples} below.

\section{Elicitation of prior parameter sets}
\label{sec:elicitation}

To represent expert knowledge on component failure times
through bounds for $\ykz$ and $\nkz$,
one can refer to the interpretations as given in Section~\ref{sec:weibull}:
$\ykz$ is the prior expected value of $\lambda_k$,
where $\lambda_k$ is linked to expected component lifetimes through \eqref{eq:lambdainterpret}.
$\nkz$ can be seen as pseudocount, indicating how strong expert knowledge is trusted
in comparison to a sample of size $n$.
Crucially, the approach allows the expert to give ranges $[\ykzl, \ykzu]$ and $[\nkzl, \nkzu]$
instead of requiring a precise answer.

It is also possible to directly link $\nkz$ and $\ykz$
to observed lifetimes using a prior predictive distribution.
Dropping the component index $k$ for ease of notation, this is given by
\begin{linenomath*}
\begin{align}
f(t\mid \nz, \yz)
 &= \int f(t\mid \lambda) f(\lambda\mid\nz,\yz) \dd \lambda \nonumber\\
 &= \beta\, t^{\beta - 1}\, (\nz + 1) \frac{(\nz \yz)^{\nz + 1}}{(\nz \yz + t^\beta)^{\nz + 2}} \,.
\label{eq:tpriopred}
\end{align}
\end{linenomath*}
Replacing the prior parameters $\nz$ and $\yz$
with their posterior counterparts $\nn$ and $\yn$ as defined in \eqref{eq:ig-update},
the effect of virtual observations on \eqref{eq:tpriopred}, or the corresponding reliability function, can be determined.
This allows to determine $\nz$ and $\yz$ through a number of `what-if' scenarios,
by asking the expert to state what (s)he would expect to learn from observing certain virtual data.

This strategy is known as pre-posterior analysis,
being first advocated by \citeN[p.~19]{1965:good}.
We recommend to check whether the effects of $\PkZ$ on the inference of interest
(this may not always be the full reliability function)
reasonably reflect an expert's beliefs before the data and in case some specific data become available,
both data agreeing with initial beliefs and surprising data.
Essentially, we advise to do an analysis like in our examples in Section~\ref{sec:examples} below,
using hypothetical data.

To elicit a meaningful prior distribution, or a set of prior distributions,
it is important to ask questions which enable experts to stay close to their actual expertise.
\citeN{1996:coolen::cens} discussed the possibility of generalizing the usual conjugate prior distributions,
for parameters of exponential family models, by including pseudo-data which are right-censored.
If the real data set contains such values, then such generalized priors do not lead to more computational complexities,
while they can have several advantages.
In addition to providing slightly more general classes of prior distributions through an additional hyperparameter,
they may enable more realistic elicitation of expert judgements,
for example if the expert has no experience with certain components past a specific life time.
For more details we refer to \citeN{1996:coolen::cens},
it should be noted that adopting such generalized prior distributions
may also provide more flexibility for modelling the effects of prior-data conflict,
this is left as a topic for future research.

\begin{figure}
\centering
\begin{tikzpicture}
[typeM/.style={rectangle,draw,fill=black!20,thick,inner sep=0pt,minimum size=8mm}, 
 typeC/.style={rectangle,draw,fill=black!20,thick,inner sep=0pt,minimum size=8mm}, 
 typeP/.style={rectangle,draw,fill=black!20,thick,inner sep=0pt,minimum size=8mm}, 
 typeH/.style={rectangle,draw,fill=black!20,thick,inner sep=0pt,minimum size=8mm}, 
 type1/.style={rectangle,draw,fill=black!20,very thick,inner sep=0pt,minimum size=8mm},
 type2/.style={rectangle,draw,fill=black!20,very thick,inner sep=0pt,minimum size=8mm},
 type3/.style={rectangle,draw,fill=black!20,very thick,inner sep=0pt,minimum size=8mm},
 cross/.style={cross out,draw=red,very thick,minimum width=9mm, minimum height=7mm},
 hv path/.style={thick, to path={-| (\tikztotarget)}},
 vh path/.style={thick, to path={|- (\tikztotarget)}}]
\begin{scope}[xscale=1.5, yscale=1.2]
\node[typeM] (M)    at ( 0  , 0  ) {M};
\node[typeC] (C1)   at ( 1  , 1.5) {C1};
\node[typeC] (C2)   at ( 1  , 0.5) {C2};
\node[cross]        at ( 1  , 0.5) {};
\node[typeC] (C3)   at ( 1  ,-0.5) {C3};
\node[cross]        at ( 1  ,-0.5) {};
\node[typeC] (C4)   at ( 1  ,-1.5) {C4};
\node[typeP] (P1)   at ( 2  , 1.5) {P1};
\node[typeP] (P2)   at ( 2  , 0.5) {P2};
\node[cross]        at ( 2  , 0.5) {};
\node[typeP] (P3)   at ( 2  ,-0.5) {P3};
\node[cross]        at ( 2  ,-0.5) {};
\node[typeP] (P4)   at ( 2  ,-1.5) {P4};
\node[typeH] (H)    at ( 0  ,-1  ) {H};
\coordinate (start)  at (-0.7, 0);
\coordinate (startC) at ( 0.5, 0);
\coordinate (startH) at (-0.4, 0);
\coordinate (Hhop1)  at ( 0.4,-1);
\coordinate (Hhop2)  at ( 0.6,-1);
\coordinate (endP)   at ( 2.5, 0);
\coordinate (end)    at ( 2.8, 0);
\path (start)     edge[hv path] (M.west)
      (M.east)    edge[hv path] (startC)
      (startC)    edge[vh path] (C1.west)
                  edge[vh path] (C2.west)
                  edge[vh path] (C3.west)
                  edge[vh path] (C4.west)
      (C1.east)   edge[hv path] (P1.west)
      (C2.east)   edge[hv path] (P2.west)
      (C3.east)   edge[hv path] (P3.west)
      (C4.east)   edge[hv path] (P4.west)
      (endP)      edge[vh path] (P1.east)
                  edge[vh path] (P2.east)
                  edge[vh path] (P3.east)
                  edge[vh path] (P4.east)
                  edge[hv path] (end)
      (startH)    edge[vh path] (H.west)
      (H.east)    edge[hv path] (Hhop1)
      (Hhop1)     edge[thick,out=90,in=90] (Hhop2)
      (Hhop2)     edge[hv path] (P3.south)
                  edge[hv path] (P4.north);
\end{scope}
\end{tikzpicture}
\caption{Reliability block diagram for a simplified automotive brake system,
with those components marked that we assume to fail in the three scenarios.}
\label{fig:brakesys-layout}
\end{figure}
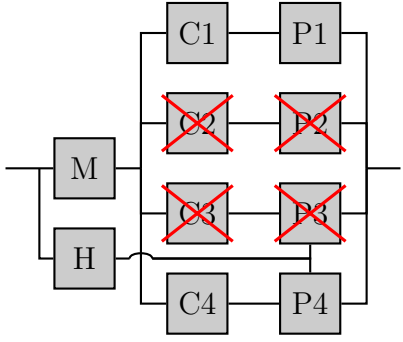

\section{Examples}
\label{sec:examples}

As illustrative example, consider a simplified automotive brake system with four types of component.
The master brake cylinder (M) activates all four wheel brake cylinders (C1 -- C4),
which in turn actuate a braking pad assembly each (P1 -- P4).
The hand brake mechanism (H) goes directly to the brake pad assemblies P3 and P4;
the car brakes when at least one brake pad assembly is actuated.
The system layout is depicted in Figure~\ref{fig:brakesys-layout},
with those components marked that we assume to fail in each of the three cases studied below.
The values for $\Phi \not\in \{0,1\}$ for the complete system are given in Table~\ref{tab:brake-survsign}.

\begin{table}
\centering
\begin{tabular}{cccclcccccl}
  \toprule
M & H & C & P & $\Phi$ & \quad & M & H & C & P & $\Phi$\\ 
  \midrule
1 & 0 & 1 & 1 & 0.25 & & 1 & 0 & 2 & 1 & 0.50 \\ 
1 & 0 & 1 & 2 & 0.50 & & 1 & 0 & 2 & 2 & 0.83 \\ 
1 & 0 & 1 & 3 & 0.75 & & 1 & 0 & 3 & 1 & 0.75 \\ 
0 & 1 & 0 & 1 & 0.50 & & 1 & 1 & 0 & 1 & 0.50 \\ 
0 & 1 & 0 & 2 & 0.83 & & 1 & 1 & 0 & 2 & 0.83 \\ 
0 & 1 & 1 & 1 & 0.62 & & 1 & 1 & 1 & 1 & 0.62 \\ 
0 & 1 & 1 & 2 & 0.92 & & 1 & 1 & 1 & 2 & 0.92 \\ 
0 & 1 & 2 & 1 & 0.75 & & 1 & 1 & 2 & 1 & 0.75 \\ 
0 & 1 & 2 & 2 & 0.97 & & 1 & 1 & 2 & 2 & 0.97 \\ 
0 & 1 & 3 & 1 & 0.88 & & 1 & 1 & 3 & 1 & 0.88 \\ 
   \bottomrule
\end{tabular}
\caption{Survival signature values $\not\in \{0,1\}$ for the simplified automotive brake system depicted in Figure~\ref{fig:brakesys-layout}.}
\label{tab:brake-survsign}
\end{table}

A fixed prior setting, described in Section~\ref{sec:ex-prior},
will be combined with three different data scenarios,
where one observes failure times in accordance with prior expectations in the first case (Section~\ref{sec:ex-case1}),
surprisingly early failures in the second case (Section~\ref{sec:ex-case2}), and
surprisingly late failures in the third case (Section~\ref{sec:ex-case3}).
In each case, it is assumed that C2, C3, P2 and P3 fail, only the failure times are varied,
investigating the effect of learning about these failures on the component level.
Effects on posterior reliability bounds for the running system
are then discussed for all three cases in Section~\ref{sec:ex-sysrel}.

\subsection{Prior assumptions}
\label{sec:ex-prior}

The prior assumptions, which one can imagine to be determined by an expert,
or by a combination of expert knowledge and component test data,
are given by the prior parameter sets $\PkZ$, $k=\text{M}, \text{H}, \text{C}, \text{P}$,
as described in Table~\ref{tab:priorparamsets}.
There, $\El[T_i^k]$ and $\Eu[T_i^k]$ give the lower and upper bound for expected component lifetimes, respectively,
which then have been transformed to bounds for the scale parameter using \eqref{eq:lambdainterpret},
resulting in $\ykzl$ and $\ykzu$.
For example, according to the expert, the mean time to failure for component type M is between 5 and 8 time units,
leading to $\ykzlfun{\text{M}} = 75.4$ and $\ykzufun{\text{M}} = 244.1$,
and the expert considers his knowledge on these expected lifetime bounds
as having the strength of at least 2 and at most 5 observations.  
\begin{table}
\centering
\begin{tabular}{crrrrrrr}
  \toprule
$k$ & $\beta_k$ & $\El[T_i^k]$ & $\Eu[T_i^k]$ & $\ykzl$ & $\ykzu$ & $\nkzl$ & $\nkzu$ \\
  \midrule
M & $2.5$ & $5$\rule{1.5ex}{0ex} & $ 8$\rule{1.5ex}{0ex} & $75.4$ & $244.1$ & $2$\rule{1ex}{0ex} & $ 5$\rule{1ex}{0ex} \\
H & $1.2$ & $2$\rule{1.5ex}{0ex} & $20$\rule{1.5ex}{0ex} & $ 2.5$ & $ 39.2$ & $1$\rule{1ex}{0ex} & $10$\rule{1ex}{0ex} \\
C & $2  $ & $8$\rule{1.5ex}{0ex} & $10$\rule{1.5ex}{0ex} & $81.5$ & $127.3$ & $1$\rule{1ex}{0ex} & $ 5$\rule{1ex}{0ex} \\
P & $1.5$ & $3$\rule{1.5ex}{0ex} & $ 4$\rule{1.5ex}{0ex} & $ 6.1$ & $  9.3$ & $1$\rule{1ex}{0ex} & $10$\rule{1ex}{0ex} \\
  \bottomrule
\end{tabular}
\caption{Prior parameter sets for the four component types.}
\label{tab:priorparamsets}
\end{table}

\begin{figure}
\includegraphics[width=\textwidth]{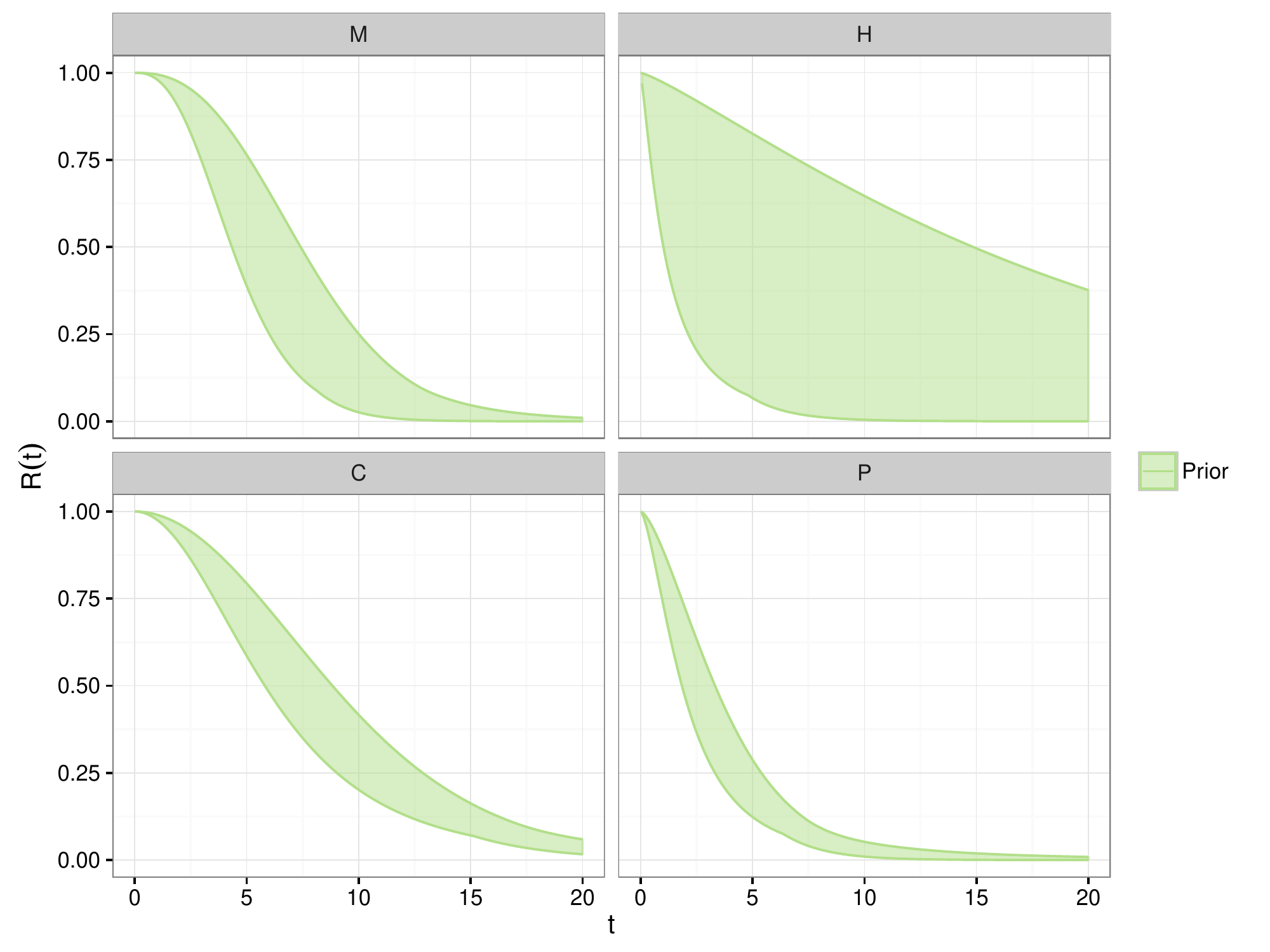}
\caption{Sets of prior predictive reliability functions for the four component types,
illustrating the choice of prior parameter sets $\PkZ$, $k=\text{M}, \text{H}, \text{C}, \text{P}$.}
\label{fig:brake-comppriors}
\end{figure}

These prior assumptions for the four component types are visualized in Figure~\ref{fig:brake-comppriors},
showing the sets of reliability functions corresponding to the prior predictive density \eqref{eq:tpriopred}.
The figure thus displays the bounds for the probability that a single component,
having been put under risk at time $0$, will have failed by time $t$.
The top left graph in Figure~\ref{fig:brake-sysrels} shows what the prior assumptions on components
signify for the system, depicting the prior bounds for the system reliability 
on a scale of time elapsed since system startup.
For example, the prior probability of the system to survive until time $10$ is between $0.03$\% and $6.91$\%.

\subsection{Case 1: failure times as expected}
\label{sec:ex-case1}

\begin{figure}
\includegraphics[width=\textwidth]{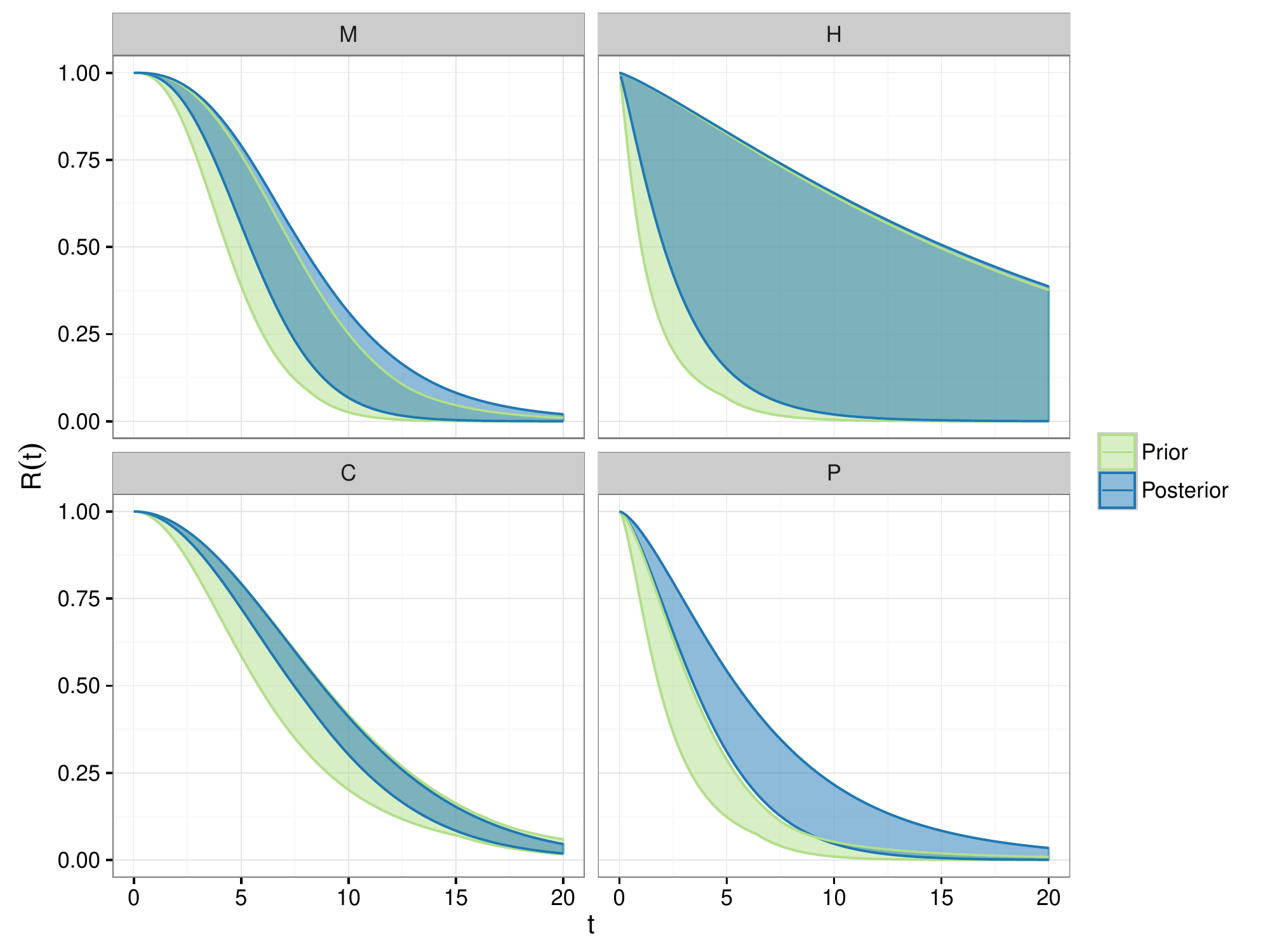}
\caption{Sets of posterior predictive reliability functions for the four component types
for observations in line with prior expectations (case~1).}
\label{fig:comppost-1}
\end{figure}

In the first case, we observe $t_1^\text{C} = 6$, $t_2^\text{C} = 7$, $t_1^\text{P} = 3$, $t_2^\text{P} = 4$,
and observe the running system until $\tnow = 8$,
i.e., $t_1^\text{M} = t_1^\text{H} = t_3^\text{C} = t_4^\text{C} = t_3^\text{P} = t_4^\text{P} = 8^+$.
(Note that component failure times are numbered by order, not by component number in the system layout.)
These observations correspond more or less to prior expectations,
and the corresponding posterior predictive component distributions are given in Figure~\ref{fig:comppost-1}.
In analogue to Figure~\ref{fig:brake-comppriors},
Figure~\ref{fig:comppost-1} displays the bounds for the probability that a single component,
having been put under risk at time $0$, will have failed by time $t$,
after having seen these (partly censored) observations.
For easy comparisons, Figure~\ref{fig:comppost-1} also contains the prior bounds from Figure~\ref{fig:brake-comppriors}.

We see that the graphs for M and H do not change dramatically,
as there is only one component of each in the system to learn from.
For C, the bounds have considerably narrowed,
showing the effect of having seen the four observations
$t_1^\text{C} = 6$, $t_2^\text{C} = 7$, $t_3^\text{C} = t_4^\text{C} = 8^+$.
The bounds for P have not narrowed as much;
this is due to the two right-censored observations $t_3^\text{P} = t_4^\text{P} = 8^+$;
from the viewpoint of the prior, surviving past time 8 is already quite unusual.

\subsection{Case 2: surprisingly early failure times}
\label{sec:ex-case2}

\begin{figure}
\includegraphics[width=\textwidth]{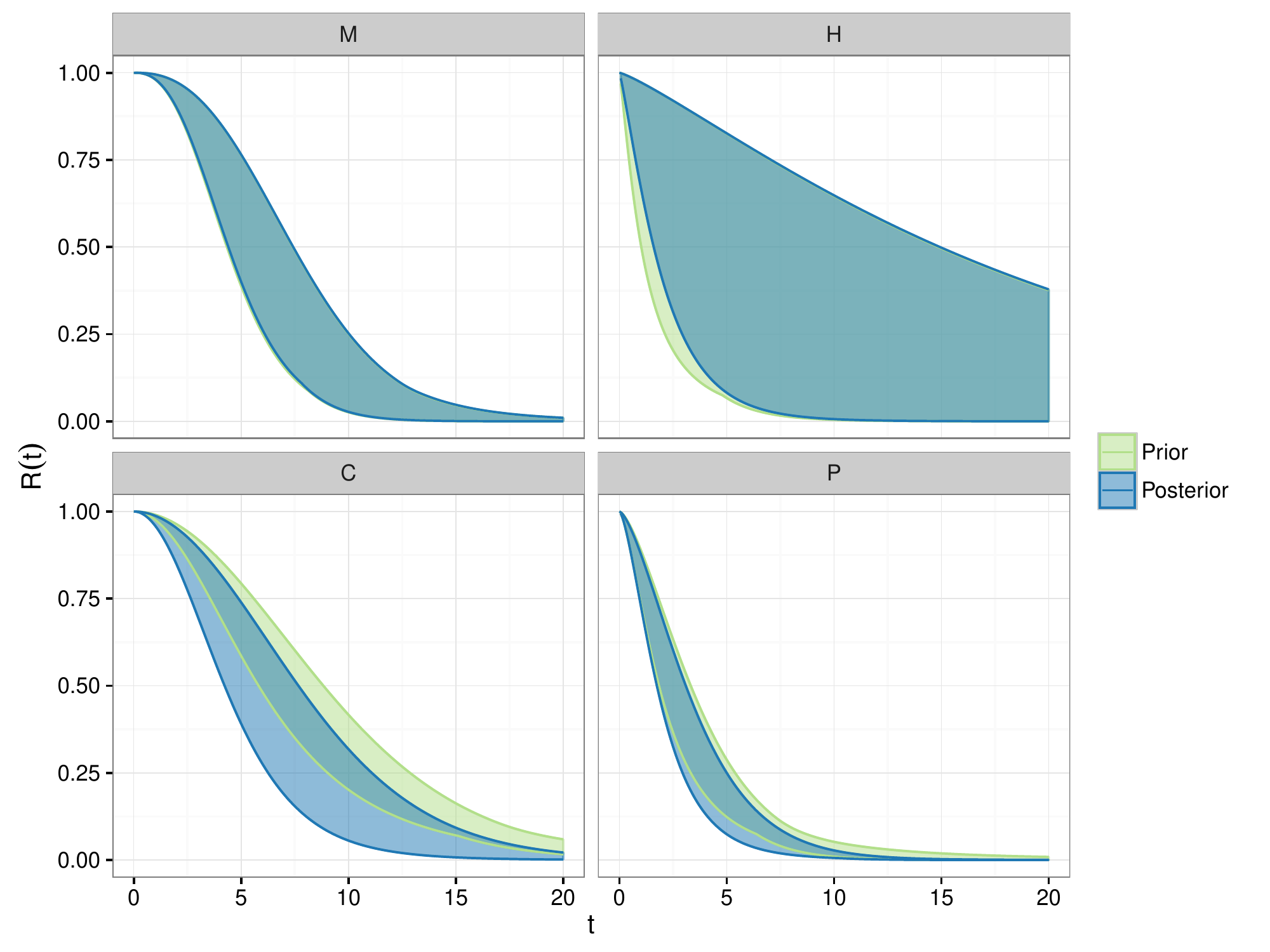}
\caption{Sets of posterior predictive reliability functions for the four component types
for surprisingly early failures (case~2).}
\label{fig:comppost-2}
\end{figure}

For the second case, with $t_1^\text{C} = 1$, $t_2^\text{C} = 2$, $t_1^\text{P} = 0.25$, $t_2^\text{P} = 0.5$
and $\tnow = 2$ (so $t_1^\text{M} = t_1^\text{H} = t_3^\text{C} = t_4^\text{C} = t_3^\text{P} = t_4^\text{P} = 2^+$),
we assume to observe surprisingly early failures;
the corresponding posterior predictive component distributions are given in Figure~\ref{fig:comppost-2}.
Having observed the system only until $t = 2$, the data are not very informative,
such that prior and posterior predictive reliability bounds are very similar.
For C, the effect of the early failures is however still visible,
and posterior imprecision, i.e., the range between lower and upper posterior bound, is notably larger
as compared to prior imprecision, and substantially larger that posterior imprecision in case~1.
The effect for P is less pronounced,
mainly because observing $\vec{t}^\text{P} = (0.25, 0.5, 2^+, 2^+)$ for P is less extreme
as observing $\vec{t}^\text{C} = (1, 2, 2^+, 2^+)$ for C.

\subsection{Case 3: surprisingly late failure times}
\label{sec:ex-case3}

\begin{figure}
\includegraphics[width=\textwidth]{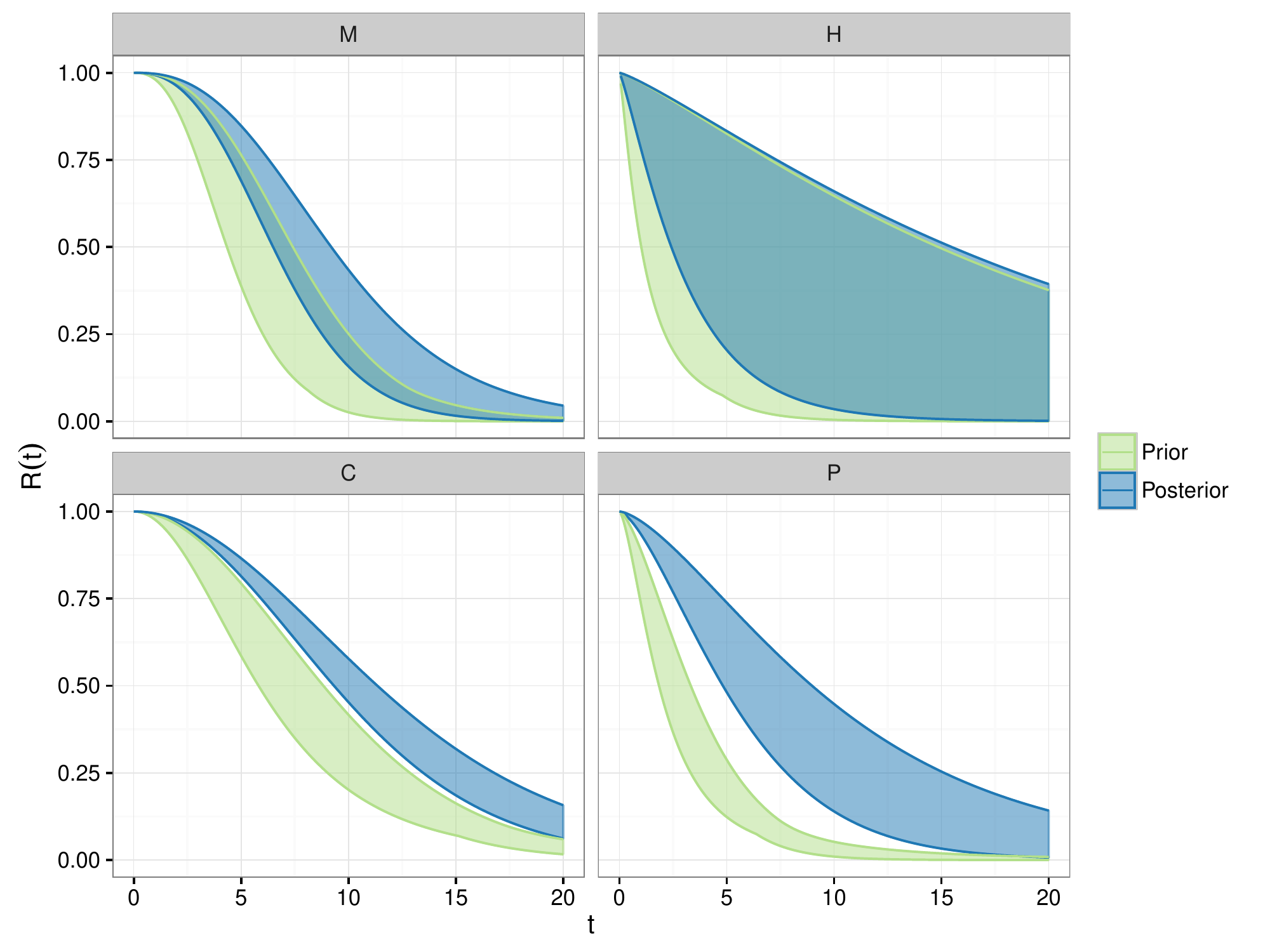}
\caption{Sets of posterior predictive reliability functions for the four component types
for surprisingly late failures (case~3).}
\label{fig:comppost-3}
\end{figure}

For the third case, we assume to observe surprisingly late failures,
namely $t_1^\text{C} = 11$, $t_2^\text{C} = 12$, $t_1^\text{P} = 8$, $t_2^\text{P} = 9$,
and $\tnow = 12$ (so $t_1^\text{M} = t_1^\text{H} = t_3^\text{C} = t_4^\text{C} = t_3^\text{P} = t_4^\text{P} = 12^+$);
the corresponding posterior predictive component distributions are given in Figure~\ref{fig:comppost-3}.
Having observed the system for a much longer time than in case 2,
the data contain now much more information,
resulting in considerable differences between prior and posterior bounds. 
The effect of these surprisingly late failures is most prominent for P,
with a set considerably shifted to the right and having very wide posterior bounds.
The posterior set for C also indicates that,
after having seen these late failures, one expects type C components to fail much later.
This effect is also visible for M and H,
but is weaker for them as there is only one component of each in the system.

\subsection{Reliability bounds for the running system}
\label{sec:ex-sysrel}

\begin{figure}
\includegraphics[width=\textwidth]{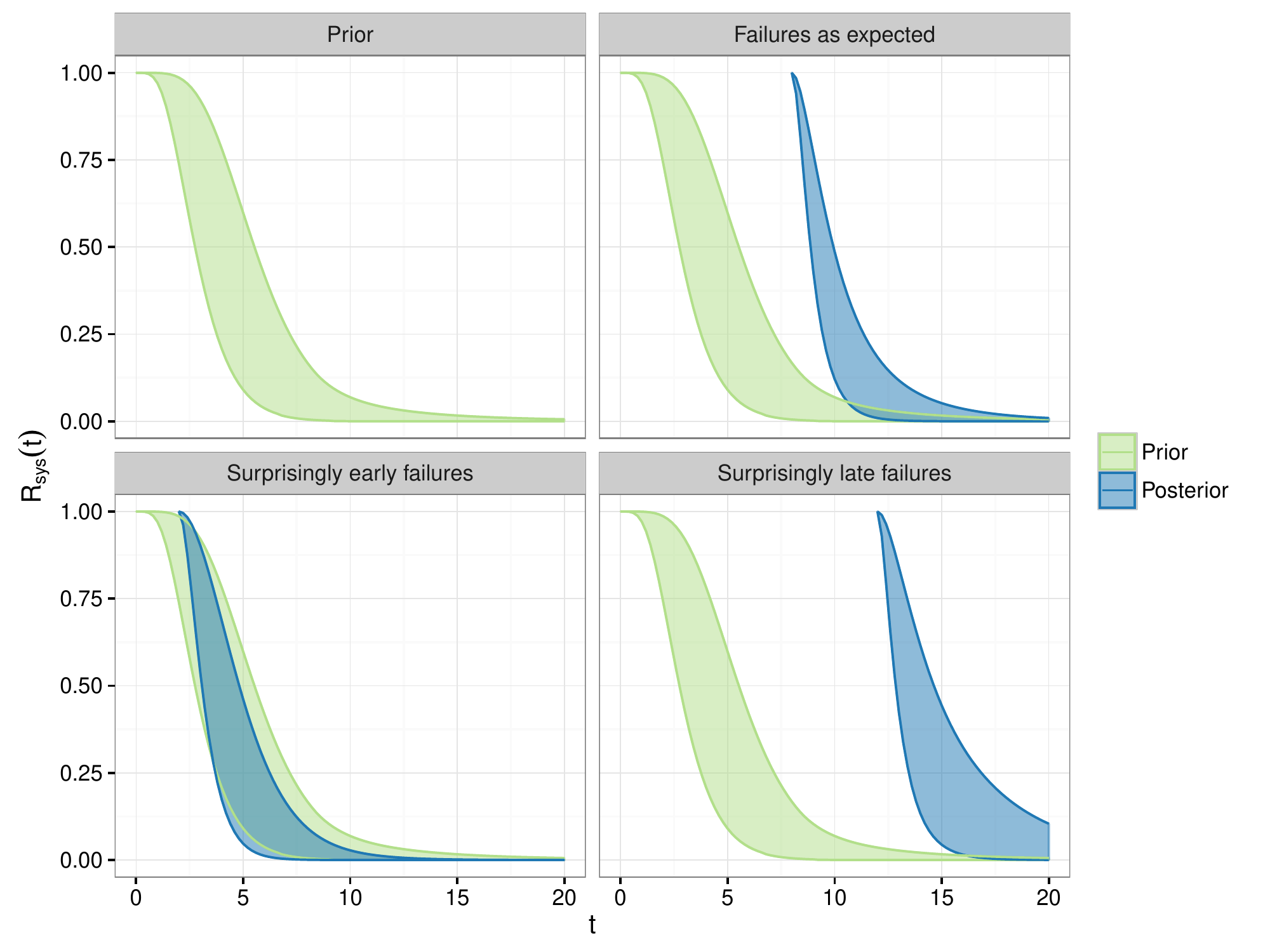}
\caption{Sets of prior and posterior system reliability functions for the three cases
on a time showing time elapsed since system startup.}
\label{fig:brake-sysrels}
\end{figure}
Figure~\ref{fig:brake-sysrels} depicts the set of prior system reliability functions,
together with the sets of posterior system reliability function for the three cases,
on a scale of elapsed time since system startup.
Due to this time scale, posterior system reliability is 1 at $\tnow$,
as it is known that the system has survived until $\tnow = 8, 2, 12$ in case 1,2,3, respectively.
For all three cases, the posterior bounds drop faster after $\tnow$ than the prior bounds drop after $t = 0$
since the components in the system have aged until $\tnow$ and so are expected to fail sooner.
In the case of surprisingly early failures, posterior bounds are mostly within prior bounds,
this is due to $\tnow$ being close to $0$ and weakly informative data in this scenario;
the posterior bounds are nevertheless wider than those for case~1;
posterior bounds are widest for case~3.
\begin{figure}
\includegraphics[width=\textwidth]{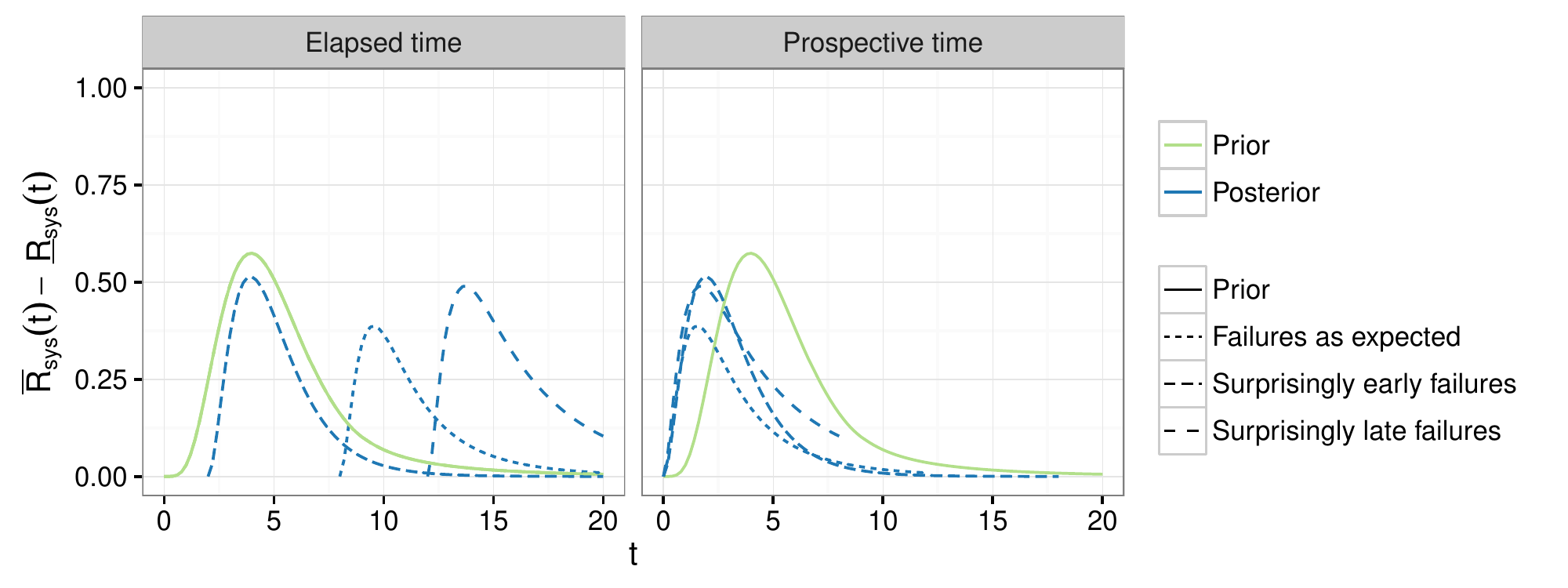}
\caption{Imprecision of prior and posterior system reliability sets.}
\label{fig:brake-sysrels-imprecision}
\end{figure}
This most visible in Figure~\ref{fig:brake-sysrels-imprecision},
which shows $\uRsys(t) - \lRsys(t)$, the difference between upper and lower bound
of prior and posterior system reliability.
The left panel shows imprecision on the scale of elapsed time since system startup like in Figure~\ref{fig:brake-sysrels};
the right panel shows imprecision on the scale of prospective time instead,
indicating how far in the future periods are for which estimation of system reliability is most uncertain.
Posterior imprecision is indeed considerably lower in case~1,
where failure times were more or less like expected.
On the prospective timescale, one can see that periods of heightened uncertainty are closer to the present for the posteriors,
while uncertainty is considerably reduced for periods further in the future.

\newpage
\section{Concluding Remarks}
\label{sec:concluding}

In this paper we presented a robust Bayesian approach to reliability estimation for systems of arbitrary layout,
and showed how the use of sets of prior distributions results in increased imprecision,
i.e., more cautious probability statements, in case of prior-data conflict (cases~2 and 3 in Section~\ref{sec:examples}),
while giving more precise reliability bounds when prior and data are in agreement (case~1 in Section~\ref{sec:examples}).
The parameters through which prior information is encoded have a clear interpretation
and are thus easily elicited, either directly or with help of the prior predictive \eqref{eq:tpriopred}.
Calculation of lower and upper predictive system reliability bounds is tractable,
requiring only a simple $K$-dimensional box-constrained optimization in Equation~\eqref{eq:sysrel-optim-n0}.

We think that increased imprecision is an appropriate tool for mirroring prior-data conflict
when considering sets of priors as is done in both the robust Bayesian and imprecise probability framework.
We want to emphazise, however, that this tool may be useful already
for just highlighting `conflict' between multiple information sources,
and that we do not think that the resulting set of posteriors,
although it can form a meaningful basis, must necessarily be used
for all consequential inferences, as a strict Bayesian would posit.
We believe an analyst is free to reconsider any aspect of a model
(of which the choice of prior can be seen to form a part) after seeing the data,
and so may use our method only for becoming aware of a conflict between prior and data.

The employed robust Bayesian setting provides many further modelling opportunities
beyond the explicit reaction to prior-data conflict.
These opportunities have not yet been explored and
provide a wide field for further research.
An example is our currently ongoing investigation into
extending the present model to allow also for an appropriate reflection of very strong agreement between prior and data
\cite{ipmu2016}.

There are many aspects to further develop in our analysis and modeling.
The general approaches used in this paper,
namely the use of sets of conjugate priors for component lifetime models
and the survival signature to calculate the system reliability,
can be used with other parametric component lifetime distributions
that form a canonical exponential family,
since for such distributions, a canonical conjugate prior
using the same canonical parameters $\nz$ and $\yz$ can be constructed,
for details see, e.g., \citeN[p.~202 and 272f]{2000:bernardosmith}, or \citeN[p.~8]{diss}.
Likewise straightforward to implement is, e.g., the analysis of the effect of replacing failed components
in the system on system reliability bounds. 
Criteria for the trade-off between the cost of replacement and the gain in reliability
would have to be adapted for the interval output of our model,
leading to very interesting research questions.

To estimate the shape parameters $\beta_k$ together with the scale parameters $\lambda_k$,
one could follow standard Bayesian approaches and use a finite discrete distribution for $\beta_k$.
Developing this together with suitable sets of priors for $\lambda_k$,
in particular to show the effect of prior-data conflict, is another interesting challenge for future research.

Another further important aspect in system reliability we have not accounted for yet
is the possibility of common-cause failures,
i.e., failure events where several components fail at the same time due to a shared or common root cause.
This could be done by combining the common-cause failure model approaches
of \citeN{Troffaes2014a} and \citeN{2015:coolen-commoncause}.

On a more abstract level, the choice for the set of prior parameters
(generating the set of prior component failure distributions) as $[\nkzl, \nkzu] \times [\ykzl, \ykzu]$
has the advantage of allowing for easy elicitation and tractable inferences,
but it may not be suitable to reflect certain kinds of prior knowledge.
Also, as studied in \cite{Walter2011a} and \cite[\S 3.1]{diss},
the shape of the prior parameter set has a crucial influence on model behaviour
like the severity of prior-data conflict reaction.
As noted in \cite[pp.~66f]{diss},
more general prior parameter set shapes are possible in principle,
but may be more difficult to elicit and make calculations more complex.

\section*{Acknowledgements}

Gero Walter was supported by the DINALOG project CAMPI
(``Coordinated Advanced Maintenance and Logistics Planning for the Process Industries'').

%
%
%
\appendix\label{section:references}
%
%
\bibliography{refs}

\begin{thebibliography}{}

\bibitem[\protect\citeauthoryear{}{Ahmadzadeh and
  Lundberg}{2014}]{2014:rul-review}
Ahmadzadeh, F. and Lundberg, J. (2014).
\newblock ``Remaining useful life estimation: review.''\ {\em International
  Journal of System Assurance Engineering and Management}, 5, 461--474.

\bibitem[\protect\citeauthoryear{}{Augustin et~al.}{2014}]{itip}
Augustin, T., Coolen, F. P.~A., de~Cooman, G., and Troffaes, M. C.~M. (2014).
\newblock {\em Introduction to Imprecise Probabilities}.
\newblock Wiley, New York.

\bibitem[\protect\citeauthoryear{}{Barlow and Proschan}{1975}]{BP75}
Barlow, R. and Proschan, F. (1975).
\newblock {\em Statistical Theory of Reliability and Life Testing}.
\newblock Holt, Rinehart and Winston, Inc., New York.

\bibitem[\protect\citeauthoryear{}{Berger et~al.}{1994}]{1994:berger}
Berger, J., Moreno, E., Pericchi, L., Bayarri, M., Bernardo, J.~M., Cano, J.,
  De~la Horra, J., Mart{\'\i}n, J., R{\'\i}os~Ins{\'u}a, D., Betr{\`o}, B.,
  Dasgupta, A., Gustafson, P., Wasserman, L., Kadane, J., Srinivasan, C.,
  Lavine, M., O'Hagan, A., Polasek, W., Robert, C., Goutis, C., Ruggeri, F.,
  Salinetti, G., and Sivaganesan, S. (1994).
\newblock ``An overview of robust {B}ayesian analysis.''\ {\em TEST}, 3,
  5--124.

\bibitem[\protect\citeauthoryear{}{Bernardo and
  Smith}{2000}]{2000:bernardosmith}
Bernardo, J. and Smith, A. (2000).
\newblock {\em Bayesian Theory}.
\newblock Wiley, Chichester.

\bibitem[\protect\citeauthoryear{}{Coolen}{1996}]{1996:coolen::cens}
Coolen, F. P.~A. (1996).
\newblock ``On {B}ayesian reliability analysis with informative priors and
  censoring.''\ {\em Reliability Engineering \& System Safety}, 53, 91--98.

\bibitem[\protect\citeauthoryear{}{Coolen and
  Coolen-Maturi}{2012}]{2012:survsign}
Coolen, F. P.~A. and Coolen-Maturi, T. (2012).
\newblock ``Generalizing the signature to systems with multiple types of
  components.''\ {\em Complex Systems and Dependability}, W. Zamojski, J.
  Mazurkiewicz, J. Sugier, T. Walkowiak, and J. Kacprzyk, eds., Vol. 170 of
  {\em Advances in Intelligent and Soft Computing}, Springer,  115--130.

\bibitem[\protect\citeauthoryear{}{Coolen and
  Coolen-Maturi}{2015}]{2015:coolen-commoncause}
Coolen, F. P.~A. and Coolen-Maturi, T. (2015).
\newblock ``Predictive inference for system reliability after common-cause
  component failures.''\ {\em Reliability Engineering \& System Safety}, 135,
  27--33.

\bibitem[\protect\citeauthoryear{}{Evans and Moshonov}{2006}]{2006:evans}
Evans, M. and Moshonov, H. (2006).
\newblock ``Checking for prior-data conflict.''\ {\em Bayesian Analysis}, 1,
  893--914.

\bibitem[\protect\citeauthoryear{}{Good}{1965}]{1965:good}
Good, I.~J. (1965).
\newblock {\em The estimation of probabilities}.
\newblock MIT Press, Cambridge (MA).

\bibitem[\protect\citeauthoryear{}{{\textsf{R} Core Team}}{2016}]{R}
{\textsf{R} Core Team} (2016).
\newblock {\em \textsf{R}: A Language and Environment for Statistical
  Computing}.
\newblock \textsf{R} Foundation for Statistical Computing, Vienna, Austria,
  $<$\url{https://www.R-project.org/}$>$.

\bibitem[\protect\citeauthoryear{}{R\'{\i}os~Insua and
  Ruggeri}{2000}]{2000:rios}
R\'{\i}os~Insua, D. and Ruggeri, F. (2000).
\newblock {\em Robust {B}ayesian Analysis}.
\newblock Springer.

\bibitem[\protect\citeauthoryear{}{Troffaes et~al.}{2014}]{Troffaes2014a}
Troffaes, M. C.~M., Walter, G., and Kelly, D.~L. (2014).
\newblock ``A robust {B}ayesian approach to modelling epistemic uncertainty in
  common-cause failure models.''\ {\em Reliability Engineering \& System
  Safety}, 125, 13--21.

\bibitem[\protect\citeauthoryear{}{Walley}{1991}]{1991:walley}
Walley, P. (1991).
\newblock {\em Statistical Reasoning with Imprecise Probabilities}.
\newblock Chapman and Hall, London.

\bibitem[\protect\citeauthoryear{}{Walter}{2013}]{diss}
Walter, G. (2013).
\newblock ``Generalized {B}ayesian inference under prior-data conflict.''\
  Ph.D. thesis, Department of Statistics, LMU Munich, Department of Statistics,
  LMU Munich, $<$\url{http://edoc.ub.uni-muenchen.de/17059/}$>$.

\bibitem[\protect\citeauthoryear{}{Walter and Augustin}{2009}]{Walter2009a}
Walter, G. and Augustin, T. (2009).
\newblock ``Imprecision and prior-data conflict in generalized {B}ayesian
  inference.''\ {\em Journal of Statistical Theory and Practice}, 3, 255--271.

\bibitem[\protect\citeauthoryear{}{Walter et~al.}{2011}]{Walter2011a}
Walter, G., Augustin, T., and Coolen, F.~P. (2011).
\newblock ``On prior-data conflict in predictive {B}ernoulli inferences.''\
  {\em ISIPTA'11: Proceedings of the Seventh International Symposium on
  Imprecise Probabilities: Theories and Applications}, F.~P. Coolen, G.
  deCooman, T. Fetz, and M. Oberguggenberger, eds., SIPTA,  391--400,
  $<$\url{http://www.sipta.org/isipta11/proceedings/046.html}$>$.

\bibitem[\protect\citeauthoryear{}{Walter and Coolen}{2016}]{ipmu2016}
Walter, G. and Coolen, F. P.~A. (2016).
\newblock ``Sets of priors reflecting prior-data conflict and agreement.''\
  {\em Information Processing and Management of Uncertainty in Knowledge-Based
  Systems: 16th International Conference, IPMU 2016, Eindhoven, The
  Netherlands, June 20-24, 2016, Proceedings, Part I}, P.~J. Carvalho, M.-J.
  Lesot, U. Kaymak, S. Vieira, B. Bouchon-Meunier, and R.~R. Yager, eds., Cham,
  Springer International Publishing,  153--164,
  $<$\url{http://dx.doi.org/10.1007/978-3-319-40596-4_14}$>$.

\bibitem[\protect\citeauthoryear{}{Walter et~al.}{2015}]{2015:walter}
Walter, G., Graham, A., and Coolen, F. P.~A. (2015).
\newblock ``Robust {B}ayesian estimation of system reliability for scarce and
  surprising data.''\ {\em Safety and Reliability of Complex Engineered
  Systems: ESREL 2015}, L. Podofillini, B. Sudret, B. Stojadinovi\'{c}, E. Zio,
  and W. Kr\"{o}ger, eds., CRC Press,  1991--1998.

\end{thebibliography}
%
%
%
%

\end{document}